\documentclass[11pt]{article}\topmargin=-0.7cm \textheight 222mm \textwidth 15.8cm\oddsidemargin=2mm
\newcommand{\comm}[1]{}
\def\xxxonly{\comm}
\def\noxxx{}
\usepackage{amssymb,epsfig}
\usepackage{amsthm}
\usepackage[numbers]{natbib}
\def\citet{\cite}
\usepackage{etoolbox}
\newtheorem{theorem}{Theorem}
\newtheorem{lemma}{Lemma}
\newtheorem{proposition}{Proposition}
\newtheorem{corollary}{Corollary}
\newtheorem{definition}{Definition}
\newtheorem{remark}{Remark}

\def\defi{\stackrel{{\scriptscriptstyle \Delta}}{=}}

\def\OO{{\scriptscriptstyle O}}

\def\d{\delta}
\def\o{\omega}
\def\O{\Omega}

\def\Y{{\cal Y}}

\def\w{\widehat}
\def\Ind{{\,\rm Ind\,}}
\def\Ind{{\mathbb{I}}}

\def\const{{\rm const\,}}

\def\R{{\bf R}}

\def\Z{{\cal Z}}

\def\s{\delta}

\def\C{{\bf C}}

\def\ww{\widetilde}
\def\X{{\cal X}}
\def\t{\theta}
\def\oo{\bar}
\def\s{\sigma}
\def\D{{\Delta}}

\def\A{{\cal A}}
\def\M{{\cal M}}
\def\B{{\cal B}}

\newcommand{\be}{\begin{equation}}
\newcommand{\ee}{\end{equation}}
\newcommand{\bd}{\begin{displaymath}}
\newcommand{\ed}{\end{displaymath}}
\newcommand{\ba}{\begin{array}{ll}}
\newcommand{\ea}{\end{array}}
\newcommand{\baa}{\begin{eqnarray}}
\newcommand{\eaa}{\end{eqnarray}}
\newcommand{\baaa}{\begin{eqnarray*}}
\newcommand{\eaaa}{\end{eqnarray*}}
\font\sm=cmr10

\def\oo{\bar}

\def\A{{\cal A}({\cal T})}

\def\BL{\b_{\scriptscriptstyle min}}
\def\BL{{\scriptscriptstyle BL}}

\def\sinc{{\rm sinc\,}}
\def\ew{\left(e^{i\o}\right)}
\def\T{{\mathbb{T}}}
\def\ZZ{{\mathbb{Z}}}

\def\TT{{\D}}

\def\M{M}
\def\D{D}
\def\D{{\ZZ\backslash\{s\}}}
\def\D{{\ZZ\setminus M_s}}
\def\A{{\rm A}}
\def\B{{\rm B}}
\def\OO{\scriptscriptstyle{\O}}
\def\XN{\ell_2^{\BL,\OO}}
\def\XNL{\ell_2^{\BL,\OO}(-\infty,0)}
\def\XNL{\ell_2^{\BL,\OO}(\D)}
\def\BN{{\mathbb{B}}}
\def\om{\o_0}
\def\eww{\left(e^{\om}\right)}

\def\ellz{\ell_{1,\om,0}}
\def\elll{\ell_{1,\om,\s}}
\def\ellz{\X_{0}}
\def\elll{\X_{\s}}
\date{Submitted  April 24, 2016. Revised January 4, 2017}
\title{On exact and optimal recovering  of missing values for sequences}
\author{
Nikolai Dokuchaev \\
{\sm  Department of Mathematics \& Statistics, Curtin
University,} {\sm  GPO Box U1987, Perth 6845,}\\ {\sm Western Australia, Australia}}
\begin{document}
\def\break{}%
\def\brea{}
\def\breakk{}
\def\brea{\nonumber\\ }\def\breakk{\nonumber\\&&} 
\vspace{-1cm}
\maketitle
\let\thefootnote\relax\footnote{Accepted to {\em Signal Processing}}
 \begin{abstract} The paper studies recoverability of missing values for  sequences in a pathwise setting without probabilistic assumptions. This setting  is oriented on a situation where the underlying sequence is considered as a sole sequence rather than a member of an ensemble with known statistical properties. Sufficient  conditions of recoverability  are obtained; it is shown that sequences are recoverable if there is a certain degree of degeneracy of  the Z-transforms. We found that,  in some cases, this degree can be measured as the number of the derivatives of Z-transform vanishing at  a point. For processes with non-degenerate Z-transform, an optimal recovering based on the projection on a set of  recoverable sequences is suggested.   Some robustness of the solution  with respect to noise contamination and truncation  is established.
\par
{\bf Key words}: data recovery, discrete time, sampling theorem, band-limited interpolation.
\end{abstract}
\vspace{-0.5cm}
\section{Introduction} The paper studies optimal recovering of missing values
 for sequences, or discrete time deterministic processes. This important problem was studied intensively.
The classical results for stationary stochastic processes
with the  spectral density $\phi$ is that
a single missing   value is recoverable  with zero
error
 if and only if \baa
\int_{-\pi}^\pi \phi(\o)^{-1} d\o=-\infty.\label{Km} \eaa
(Kolmogorov \cite{K},
Theorem 24). Stochastic stationary Gaussian processes without this property are called {\em minimal } \cite{K}. In particular, a process is recoverable
if it is ``band-limited" meaning that the spectral density is vanishing on an arc of the
unit circle $\T=\{z\in\C:\ |z|=1\}$.
This illustrates the relationship of recoverability  with the notion of bandlimitiness or its relaxed versions such as (\ref{Km}).
In particular, criterion (\ref{Km}) was extended on stable processes \citet{Peller}  and vector Gaussian processes  \citet{Pou}.

In theory, a process can be converted into a band-limited and recoverable process   with a low-pass filter.     However,
 a ideal low-pass filter cannot be applied if there are  missing values. This leads to approximation
 and optimal  estimation  of missing values.
 For the forecasting and other applications, it is common to use band-limited approximations of non-bandlimited underlying processes.
 There are many works devoted to  smoothing and sampling an based on frequency properties; see e.g.
\citet{Alem,Cai,CTao,CJR,D12a,D12b,D16,Donoho,F92,F94a,F94,K,Peller,Pou,Pou2,T}.
  \index{\citet{jerri}, \citet{PFG}, \citet{PFG1}, \citet{AU}.} 

The present paper also consider band-limited approximations. We consider approximation of an observed sequence  in $\ell_r$-norms rather than  matching  the values at
selected points. The solution is not error-free; the error can be significant if the underlying process is not band-limited.
This is different from  a setting in \xxxonly{\citet{Cai,CTao,CJR,F94,LeeF}}\noxxx{\citet{Cai,CTao,CJR,F94,LeeF}}, where
error-free recovering was considered.
Our setting is closer to the setting from  \citet{TH,Zhao}.
In \citet{TH}, optimization was considered as  minimization of the total energy  for  an approximating bandlimited process
 within a given distance from  the original process smoothed by an ideal low-pass filter.
 In \citet{Zhao},  extrapolation of a band-limited process matching a finite number of points process was considered using special Slepian's type basis in the frequency domain.

The present paper considers  optimal recovering of missing values  of sequences (discrete time processes)  based on intrinsic properties of
 sequences, in the pathwise setting, without using  probabilistic assumptions on the ensemble.
This setting targets a scenario where   a sole underlying sequence  is deemed to be unique
  and such that one cannot rely on statistics  collected from observations of other similar samples.
 To address this, we use a
pathwise optimality criterion that does not involve an expectation  on a probability space.
For this  setting,  we obtained explicit optimal estimates for  missing values of a general type processes  (Theorems \ref{Th1} and \ref{Th2}).
We identified some classes of processes  with degenerate Z-transforms
 allowing  error-free recoverability  (Corollary  \ref{corr1} and \ref{corr2}). For a special case of a single missing values,
 this gives     a condition of  error-free recoverability of sequences reminding classical
 criterion  (\ref{Km}) for stochastic processes  but based on intrinsic properties of
 sequences, in the pathwise setting (Corollary  \ref{corr2}).
 In addition,   we established  numerical stability and robustness of the method  with respect to the input errors and data truncation (Section \ref{sectR}).

\section{Some definitions and background}

Let $\ZZ$  be the set of all integers. For a set $G\subset \ZZ$ and $r\in[1,\infty]$,
we denote by $\ell_r(G)$ a Banach
space of complex valued sequences $\{x(t)\}_{t\in G}$
such that
$\|x\|_{\ell_r(G)}\defi \left(\sum_{t\in G}|x(t)|^r\right)^{1/r}<+\infty$ for $r\in[1,+\infty)$,
and $\|x\|_{r(G)}\defi \sup_{t\in G}|x(t)|<+\infty$ for $r=\infty$.

\par
For  $x\in \ell_2(\ZZ)$, we denote by $X=\Z x$ the
Z-transform  \baaa X(z)=\sum_{t=-\infty}^{\infty}x(t)z^{-t},\quad
\eaaa defined for $z\in\C$ such that the series converge. For $x\in\ell_2(\ZZ)$, the function $X\ew|_{\o\in(-\pi,\pi]}$ is defined as an element
of $L_2(-\pi,\pi)$. For $x\in\ell_1(\ZZ)$, the function $X\ew$ is defined for all $\o\in(-\pi,\pi]$ and is continuous in $\o$.

Let  $m\in\ZZ$  be given, $m\ge 0$.  For  $s\in \ZZ$, let $M_s=\{s,s+1,s+2,...,s+m\}$.

We consider data recovery problem  for input processes $x\in\ell_r$ such that
the trace  $\{x(t)\}_{t\in \D}$ represents the  available observations;
the values  $\{x(t)\}_{t\in M_s}$  are missing.

\begin{definition}\label{def1} Let $\Y\subset \ell_r$ be a class of sequences.
We say that this class is  recoverable if, for any $s\in\ZZ$,
there exists a mapping  $F:\ell_r(\D)\to\R^{m+1}$ such that
$x|_{M_s}=F\left(x|_{\D}\right)$ for all $x\in \Y$.
\end{definition}

For a sequence that does not belong to a recoverable class, it is natural to accept,  as
an approximate solution, the corresponding values of  the closest process from a preselected
recoverable class. More precisely, given observations $x|_{\D}$ and a recoverable class $\Y\subset \ell_r$,
we suggest to find  an optimal solution  $\w x\in\Y$
of the minimization problem \baa &&\hbox{Minimize}\quad  \sum_{t\in \D} |\w x(t)-x(t)|^2  \quad\breakk\hbox{over}\quad \w x\in \Y\label{min}, \eaa
and accept the trace  $\w x|_{M_s}$ as the recovered missing values $x|_{M_s}$.


\section{Recovering based on band-limited smoothing}

We assume that we are given $\O\in(0,\pi)$. Let $\XN$ be the set of all $x\in\ell_2(\ZZ)$ such that $X\ew =0$ for $|\o|>\O$ for $X=\Z x$.
We will call  sequences $x\in \XN$ {\em band-limited}.
Let $\XN(\D)$ be the subset of $\ell_2(\D)$ consisting of traces
$x|_{\D}$ for all sequences  $x\in\XN$.

\begin{proposition}\label{propU}  For
  any $x\in\XNL$, there exists a unique $\w x\in\XN$ such that $\w x(t)=x(t)$ for $t\in \D$.
\end{proposition}
\par
In a general case, where the sequence of observations $x|_\D$ does not necessarily  represents a trace of a band-limited process, we will
be using approximation described in the following lemma.
\begin{lemma}\label{lemma1} There exists a unique optimal solution  $\w x\in\XN$
of the minimization problem  (\ref{min}) with $r=2$ and $\Y=\XN$.
\end{lemma}
\par
Under the assumptions of Lemma \ref{lemma1},
there exists a unique band-limited process $\w x$ such that the trace $\w x|_{\D}$
provides an optimal approximation of
its observable trace  $x|_{\D}$.   The corresponding  trace $\w x|_{M_s}$ is uniquely defined and can
be interpreted  as the solution of the problem of optimal recovering of the missing values $x|_{M_s}$
(optimal in the sense of problem (\ref{min}) given $\O$).
In this setting, the process $\w x$ is deemed to be a smoothed  version of $x$, and the process $\eta=x-\w x$ is deemed to be an irregular noise.
This justifies acceptance of $\w x|_{M_s}$ as an estimate of missing values.
It can be noted that  the recovered values depend on the choice of $\O$; the selection of  $\O$
has to be based on some presumptions about cut-off frequencies suitable for particular applications.

Let $H(z)$ be the transfer function for an ideal low-pass filter such that $H\ew=\Ind_{[-\O,\O]}(\o)$, where
$\Ind$ denotes the indicator function. Let $h=\Z^{-1}H$;
it is known that  $h(t)=\O\,\sinc(\O t)/\pi$; we use the notation $\sinc(x)=\sin(x)/x$, and we  use notation $\circ$  for the convolution in $\ell_2(\ZZ)$.
The  definitions imply that $h\circ x\in \XN$ for any $x\in \ell_2(\ZZ)$.

Consider a matrix $\A=\{h(k-p)\}_{k=0,p=0}^{m,m}\in \R^{(m+1)\times (m+1)}$.
Let $I_{m+1}$ be the unit matrix in $ \R^{(m+1)\times (m+1)}$.
\begin{lemma}\label{lemma3} The matrix $I_{m+1}-\A$ is non-degenerate.
\end{lemma}

\begin{theorem}\label{Th1}  Let $x\in\ell_2(\ZZ)$ and $\O\in (0,\pi)$. Given observations $x|_{\D}$, the problem  (\ref{min})
 with $r=2$ and $\Y=\XN$
 has a unique optimal  solution $\w x\in\XN$ which yields an estimate of $x|_{M_s}$ defined as
 \baa
\w x(s+p)=y_p, \quad p=0,1,..., m,
\label{yy1}\eaa
where   $y=\{y_p\}_{p=0}^m\in \C^{m+1}$ is defined as \baa y=(I_{m+1}-\A)^{-1}z,
\label{yy2}\eaa
with  $z=\{z_p\}_{p=0}^m\in\C^{m+1}$ defined as
\baa
 z_p= \sum_{t\in\D}h(p-t)x(t).
\label{yy3}
\eaa
\end{theorem}
\begin{corollary}\label{corr1} For any $\O\in(0,\pi)$, the class  $\XN$ is recoverable in the sense of Definition \ref{def1}.
\end{corollary}
\begin{remark}\label{remFF} Equations (\ref{yy1})-(\ref{yy3})  applied to a band-limited process $x\in \XN$ represent  a special case of the result
\citet{F92,F94a}.
The difference is that $x$ is  Theorem \ref{Th1} and (\ref{yy1})-(\ref{yy3})  is not necessarily band-limited.
\end{remark}
\subsection*{ The case of a single missing value}
It appears that the solution for the special case of a single missing value (i.e. where $m=0$) allows a convenient explicit formula.
\begin{corollary}
\label{Th11} Let  $\O\in (0,\pi)$ and $x\in\ell_2(\ZZ)$ be given.
 Given observations $x|_{\ZZ\setminus\{s\}}$, the problem  (\ref{min})
 with $r=2$ and $\Y=\XN$
 has a unique solution $\w x\in\XN$  which yields an estimate of $x(s)$  defined as
 \baa
\w x(s)=
\frac{\O}{\pi-\O} \sum_{t\in \D} x(t) \sinc[\O (s-t)].
\label{wx1}
\eaa
This solution is optimal in the sense of problem (\ref{min}) with $m=0$, $M_s=\{s\}$, $r=2$, and $\Y=\XN$,  given $\O\in(0,\pi)$.
\end{corollary}
\begin{remark}\label{rem1} Corollary \ref{Th11} applied to a band-limited process $x_\BL\in \XN$ gives a formula \baaa
x_\BL(s)=
\frac{\O}{\pi-\O} \sum_{t\in\D} x_\BL(t) \sinc[\O (s-t)].
\label{wxBL}
\eaaa
This formula is known \citet{F92,F94a}; however,  equation (\ref{wx1}) is
Corollary \ref{Th11} is different  since  $x$ in (\ref{wx1}) is not necessarily band-limited.
\end{remark}

\section{Recovering without smoothing}
Theorem \ref{Th1} suggests to replace missing values by corresponding  values of a smoothed band-limited process.
 This process is actually different from the underlying input process;   this could  cause a loss of some information  contained in high-frequency
 components. Besides, it could be difficult to justify a particular  choice of  $\O$ in (\ref{wx1}) defining the degree of smoothing.
To overcome this, we consider below the limit case where $\O\to \pi-0$.

Again, we consider input  sequences $\{x(t)\}_{t\in \D}$ representing the observations available;
the values for $t\in M_s$ are missing.

Without a loss of generality, we assume that either $s=0$ or $m=0$.

Let $\om\in (0,\pi]$ be given. For $x\in\ell_2$, l

For $\s=(\s_0,\s_1...,\s_m)\in\R^{m+1}$ such that $\s_k\ge 0$, $k=0,1,...,m$,  let
\baaa
\elll\defi \Bigl\{x\in\ell_1:\  \sum_{t\in \ZZ} |t|^m |x(t)|<+\infty,\quad
\left| \frac{d^kX}{d\o^k} \left(e^{ i\om }\right)\right|\le\s_k,
 \\ k=0,1,...,m,\quad X=\Z x\Bigr\}.
\eaaa
Here and below  we assume, as usual,  that $d^kX/d\o^k=X$ for $k=0$.

It can be shown that, for $x\in\elll$  and $X=\Z x$, we have that the functions $\frac{d^kX\ew}{d\o^k}$ are continuous in $\o$ for $k=0,1,...,m$.

\begin{definition}\label{defD}
Let $\X_0$ be the corresponding set $\X_\s$ with $\s=0$, i.e. with $\s_p=0$ for  $p=0,1,...,m$.
We will call   $x$ degenerate of order $m$.
\end{definition}


 Let us introduce a matrix  $\B(\o)=\{b_{pk}(\o)\}_{k=0,p=0}^{m,m}\in \C^{(m+1)\times (m+1)}$ such that
\baaa
b_{pk}(\o)=[-i (s+k)]^pe^{-i\o  (s+k)},\quad \o\in (-\pi,\pi].
\eaaa
In particular, if $m=0$, then $\B(\o)=e^{-i\o s}$. If $m>0$, then, by the assumptions,  $s=0$ and $b_{pk}(\o)=(-i k)^pe^{-i\o k}$.

\begin{lemma}\label{lemmaM} For any $\o\in(-\pi,\pi]$, the matrix $\B(\o)$ is non-degenerate.
\end{lemma}

\index{ For $s=0$ it is $b_{kp}=e^{-i\o_0  k}(-i k)^p$. For $\o_0=\pi$,  $b_{kp}=(-1)^k k^p$?  or
 $B=\{b_{kp}\}_{k=s,p=0}^{s+m,m}=\{b_{kp}\}_{k=s,p=0}^{m,m}$}

\begin{theorem}\label{Th2} Let  $x\in\ell_1(\ZZ)$ be given such that $\sum_{t\in \ZZ} |t|^m |x(t)|<+\infty$.  Given observations $x|_{\D}$, the problem  (\ref{min})
 with $r=1$ and $\Y=\X_0$
 has a unique solution $\w x\in\XN$  which yields an estimate  of $x|_{M_s}$ defined as
 \baa
\w x(s+p)=y_p(\o_0), \quad p=0,1,..., m,
\label{y1}\eaa
where   $y(\o)=\{y_p(\o)\}_{p=0}^m\in \C^{m+1}$ is defined as \baa y(\o)=\B(\o)^{-1}z(\o),
\label{y2}\eaa
with  $z(\o)=\{z_p(\o)\}_{p=0}^m\in\C^{m+1}$ defined as
\baa
 z_p(\o)= -\sum_{t\in\D}(-it)^p e^{-i\o t}x(t).
\label{y3}
\eaa
\end{theorem}
\par
Under the assumptions of Theorem \ref{Th2},
there exists a unique recoverable process $\w x\in\ellz$ such that  $\w x|_{t\in \D}= x|_{t\in \D}$.
The corresponding  trace $\w x|_{M_s}$ is uniquely defined and can
be interpreted  as the solution of the problem of optimal recovering of the missing values $x|_{M_s}$ (optimal in the sense of problem (\ref{min}) for $\Y=\ellz$).
In addition, Theorem \ref{Th2}  implies that  $\X_0\neq \emptyset$ for any $m\ge 0$;
this follows from the  implication from this theorem that a sequence from $\ell_1$ can be transformed into a sequence in $\elll$ by changing its $m$ terms.

\begin{corollary}\label{corr2} The class  $\X_0$ is recoverable in the sense of Definition \ref{def1} with $r=1$ and $\Y=\ellz$.
\end{corollary}
\begin{remark}\label{rem3} By Corollary \ref{corr2} applied with $m=0$, a single missing value process $x\in\ell_1$ is recoverable if
$X\eww=0$ for $X=\Z x$; this reminds condition (\ref{Km}) for spectral density of minimal Gaussian
processes \cite{K}.
\end{remark}
\subsection*{ The case of a single missing value}
Again, the solution for the special case of a single missing value (i.e. where $m=0$ and $M_s=\{s\}$) allows a simple  explicit formula.
\begin{corollary}
\label{Th21} Let $s\in\ZZ$ and $x\in\ell_1(\ZZ)$ be given.  Given observations $x|_{\ZZ\setminus\{s\}}$, the problem  (\ref{min})
 with $r=1$ and $\Y=\X_0$
 has a unique solution $\w x\in\XN$  which yields an estimate  of $x(s)$ defined as
 \baa
\w x(s)=-\sum_{t\neq  s}e^{i\om (s- t)}x(t),
\label{wx2}
\eaa where the optimality is understood in the sense of problem (\ref{min}) with $m=0$, $M_s=\{s\}$,  $r=1$, and $\Y=\ellz$.
\end{corollary}
\begin{remark}\label{rem4} Formula (\ref{wx2}) with $\om=\pi$ has the form \baa
\w x(s)=-\sum_{t\in\D} (-1)^{t-s}x(t).
\label{wx3}
\eaa This represents the limit case of formula (\ref{wx1}), since
 \baaa
\frac{\O}{\pi-\O}  \sinc[\O (s-t)]\to -(-1)^{t-s}\quad \hbox{as}\quad \O\to \pi-0
\label{limO}
\eaaa
for all $t\neq s$.
\end{remark}

\subsection*{Optimality in the minimax sense}
It will be convenient to use mappings $\d_p:\C^{m+1}\to\C$, where $p\in\{0,1,...,m\}$,  such that
$\d_p(y)=y_p$ for a vector $y=(y_0,y_1,...,y_m)\in\C^{m+1}$.
\begin{proposition}
\label{prop1}
In addition to the optimality in the sense of problem (\ref{min}) with $\Y=\ellz$, solutions obtained  in
Theorems \ref{Th2} and Corollalry \ref{Th2} are also optimal in the following sense.
\begin{enumerate}
\item If $m=0$, then solution (\ref{wx1}) is optimal in the minimax sense such that
\baa
\sup_{x\in\elll}|\w x(s)-x(s)|\le\s_0\le \sup_{x\in\X_\s}|\ww x(s)-x(s)|
\label{opt1}
\eaa
for any  estimator $\ww x(s)=F\left(x|_{\ZZ\setminus \{s\}}\right)$, where
$F:\ell_1(\ZZ\setminus \{s\})\to\C$ is a mapping.
\item If $m\ge 0$ and $s=0$, then
solution (\ref{y1})-(\ref{y3}) is optimal in the mininax sense such that
\baa
\sup_{x\in\elll}|\d_p(\B(\om) \w \eta)|\le \s_p\le \sup_{x\in\X_\s}|\d_p(\B(\om)\ww \eta)|, \quad
\brea  p=0,1,...,m,
\label{opt}
\eaa for any estimator  $\ww x|_{M_s}=F\left(x|_\D\right)$, where
$F:\ell_1(\D)\to\C^{m+1}$ is a mapping,
$\w \eta=\{\w x(t)-x(t)\}_{t=s}^{s+m}\in\C^{m+1}$,
$\ww \eta=\{\ww x(t)-x(t)\}_{t=s}^{s+m}\in\C^{m+1}$.
\end{enumerate}
\end{proposition}
\section{Robustness with respect to noise contamination and data truncation}\label{sectR}
 Let us consider a situation where an input process $x|_\D$ is observed with an error.
 In other words, assume that we observe a process $x_\eta|_\D=x|_\D+\eta|_\D$, where $\eta$ is a noise.

For a matrix $S\in \C^{m+1}$ and $r_1,r_2\in[1,+\infty]$,  we denote by $\|S\|_{r_1,r_2}$ the operator norm of this matrix
considered as an operator $S:\C^{m+1}_{r_1}\to \C^{m+1}_{r_2}$,
where $\C^{m+1}_r$ denote the linear normed space formed as $\C^{m+1}$ provided with $\ell_r$-norm.

\begin{proposition}
\label{propR1} In the notations of Theorem \ref{Th1},
\baaa
\|\w x|_{M_s}\|_{\ell_\t(M_s)}\le \left\|(I_{m+1}-\A)^{-1}\right\|_{2,\t}\|x|_{\D}\|_{\ell_2(\D)}.
\label{estR1}\eaaa
for any $\t\in[1,+\infty]$. In particular, under the assumption of Corollary \ref{Th11},
\baaa
|\w x(s)|\le \frac{\O}{\pi-\O} \|x\|_{\ell_2(\D)}.
\eaaa \end{proposition}

\begin{proposition}
\label{propR2} In the notations of Theorem \ref{Th2},
\baaa
\|\w x|_{M_s}\|_{\ell_\t(M_s)}  \le  \left\|\B(\om)^{-1}\right\|_{\infty,\t}\sum_{t\in \D} |t|^m |x(t)|\label{estR2}\eaaa
for any $\t\in[1,+\infty]$. In particular, under the assumption of Corollary \ref{Th21},
$$
|\w x(s)|\le \|x\|_{\ell_1(\D)}.
$$  \end{proposition}
\par

Propositions \ref{propR1} and \ref{propR2}
 ensure robustness  of the data recovering with respect to noise contamination  and
truncation. This can be shown as the following.

 Let $\w x_\eta|_{M_s}$ be the sequence of corresponding values defined by (\ref{yy1})-(\ref{yy3})  or (\ref{y1})-(\ref{y3}) with $x_\eta|_\D$ as an input, and let $\w x|_{M_s}$
 be the corresponding values defined by  (\ref{yy1})-(\ref{yy3}) or   with $x|_\D$ as an input.
By Proposition \ref{propR1},
  \baa
 \|(\w x-\w  x_\eta)|_{M_s}\|_{\ell_r(M_s)}  \le \|(I_{m+1}-\A)^{-1}\|_{\rho,2} \|\eta\|_{\ell_2(\D)}
\label{r1}
\eaa
 for all $\eta|_\D\in \ell_2(\D)$.
  In particular, under the assumption of Corollary \ref{Th11}, i.e. for $m=0$ and $M_s=\{s\}$, it follows  that,  in the notations of Theorem \ref{Th1},
  \baa
  |\w x(s)-\w  x(s)|\le \frac{\O}{\pi-\O} \|\eta\|_{\ell_2(\D)}.\label{r11}\eaa
\par
 Similarly, Propositions \ref{propR2} implies that
  \baa
  |\w x(s)-\w  x_\eta(s)|\le \|z_\eta(\om)\|_{\ell_1(\D)}\label{r2}\eaa
 for all $\eta|_\D\in \ell_1(\D)$, under the assumptions of this theorem, with
 $z_\eta(p,\o)=\{z_\eta(p,\o)\}_{p=0}^m\in\C^{m+1}$ defined as
\baaa
 z_\eta(p,\o)= -\sum_{t\in\D}(-it)^p e^{-i\o t}\eta(t).
\label{y33}
\eaaa

 This demonstrates some robustness of the method  with respect to  the noise in the observations.
In particular, this ensures robustness of the estimate with respect to truncation of the input processes,
such that infinite sequences $x\in \ell_r(\D)$, $r\in\{1,2\}$,
 are replaced by truncated sequences $x_\eta(t)=x(t)\Ind_{\{|t|\le q\}}$ for $q>0$; in this case
   $\eta(t)=\Ind_{|t|> q}x(t)$. Clearly,  $\|\eta\|_{\ell_r(\D)}\to 0$ as $q\to +\infty$.
 This  overcomes principal impossibility to access infinite sequences of observations.

The experiments with sequences generated by Monte-Carlo simulation
demonstrated a good numerical stability of the  method; the results were
 quite robust with respect to deviations of input processes and truncation.

 \subsection*{On a choice between recovering formulae  (\ref{wx1}) and (\ref{wx2})}
It can be seen from (\ref{r1}) and (\ref{r2}) that recovering formula (\ref{wx2})
is less robust with respect to data truncation and the noise contamination than recovering formula (\ref{wx1}). In addition, recovering formula (\ref{wx2}) is not applicable to $x\in\ell_2(\ZZ)\setminus \ell_1(\ZZ)$.   On the other hand, application of (\ref{wx2})  does not
require to select $\O$.  In practice,  numerical  implementation
requires to replace a sequence $\{x(t)\}$ by a truncated sequence
$x(t)\Ind_{\{t:\ |t|\le q\}}$; technically, this means that both formulas could be applied. The choice   between (\ref{wx1}) and (\ref{wx2}) and of a particular $\O$  for (\ref{wx1})
should be done based on the purpose of the model. In general, a  more numerically robust result can be achieved with choice of a smaller $\O$.
\par
This can be illustrated with the following example for a  case of a single missing value. Consider a band-limited input $x\in \XN$ with a missing value $x(0)$
(i.e, $m=0$ and $s=0$, in the notations above).
In theory, application of (\ref{wx1}) with $\O$ replaced by  $\O_1\in (\O,\pi]$ produces error-free recovering, i.e. $\w x(0)=x(0)$.
However, application of (\ref{wx1}) with $\O$ replaced by $\O_2\in (0,\O_1)$ may  lead to a large error
 $\w x(0)-x(0)$.
\par
\begin{figure}[ht]
\centerline{\psfig{figure=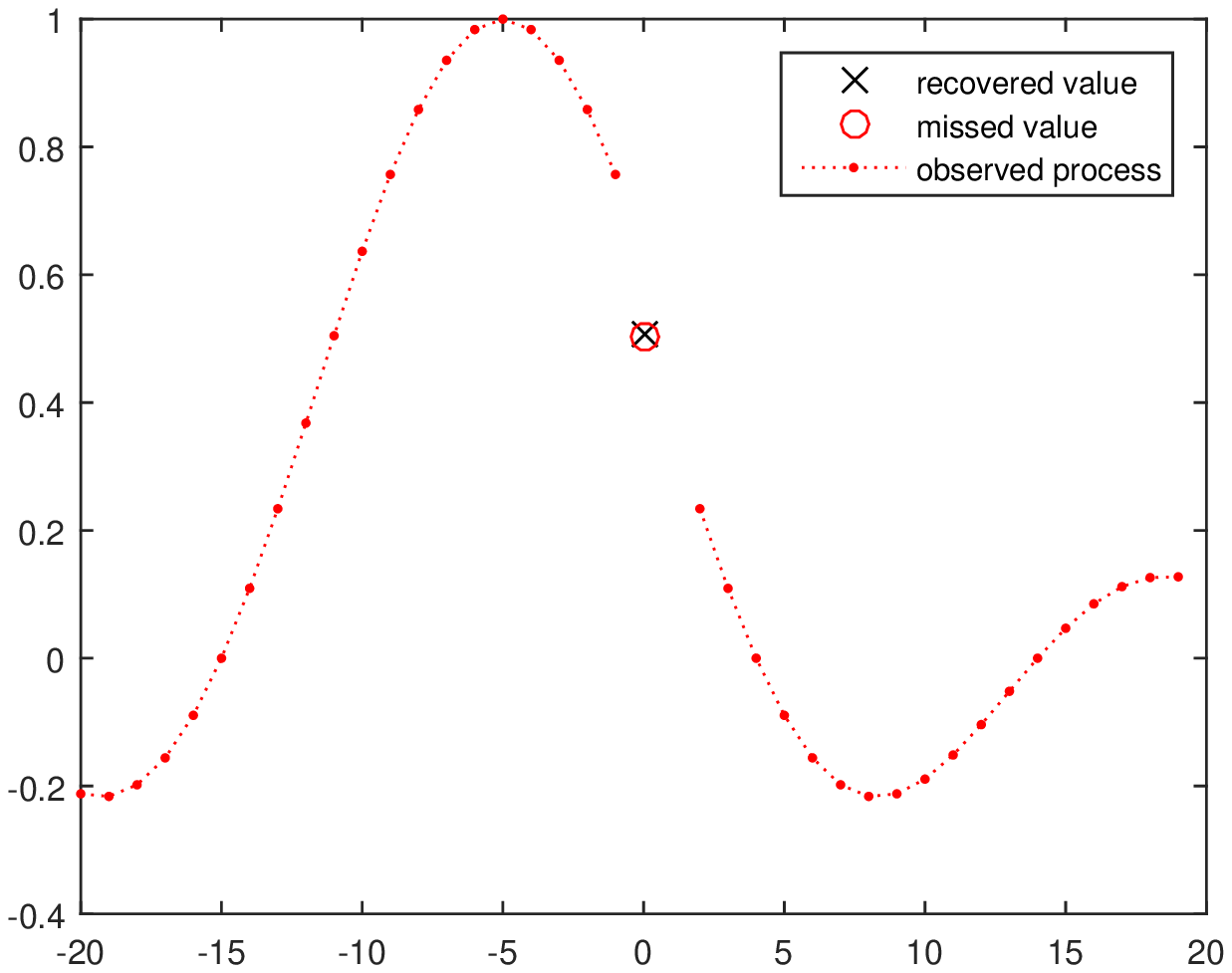,width=9cm,height=5.5cm}}
\centerline{\psfig{figure=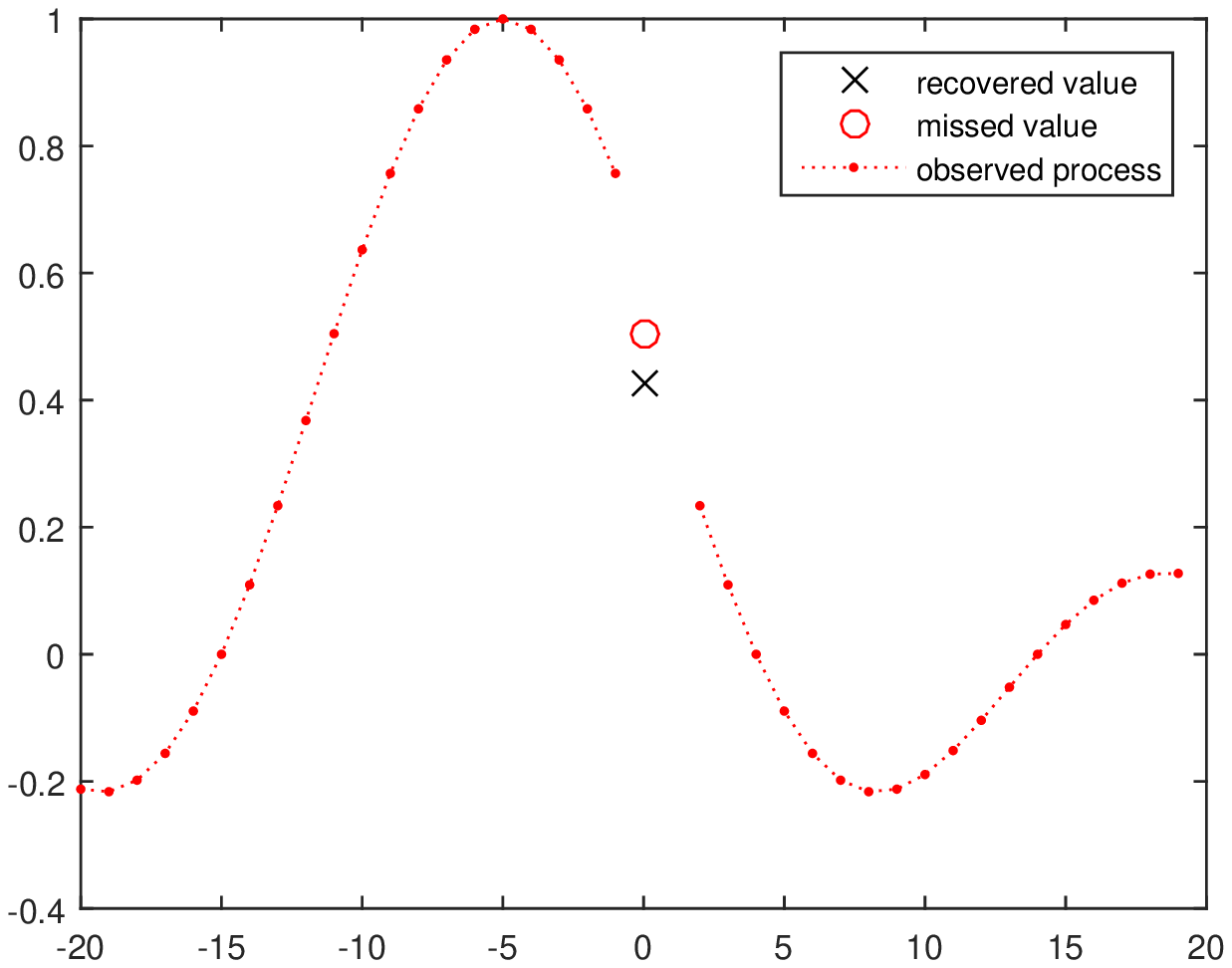,width=9cm,height=5.5cm}}
\centerline{\psfig{figure=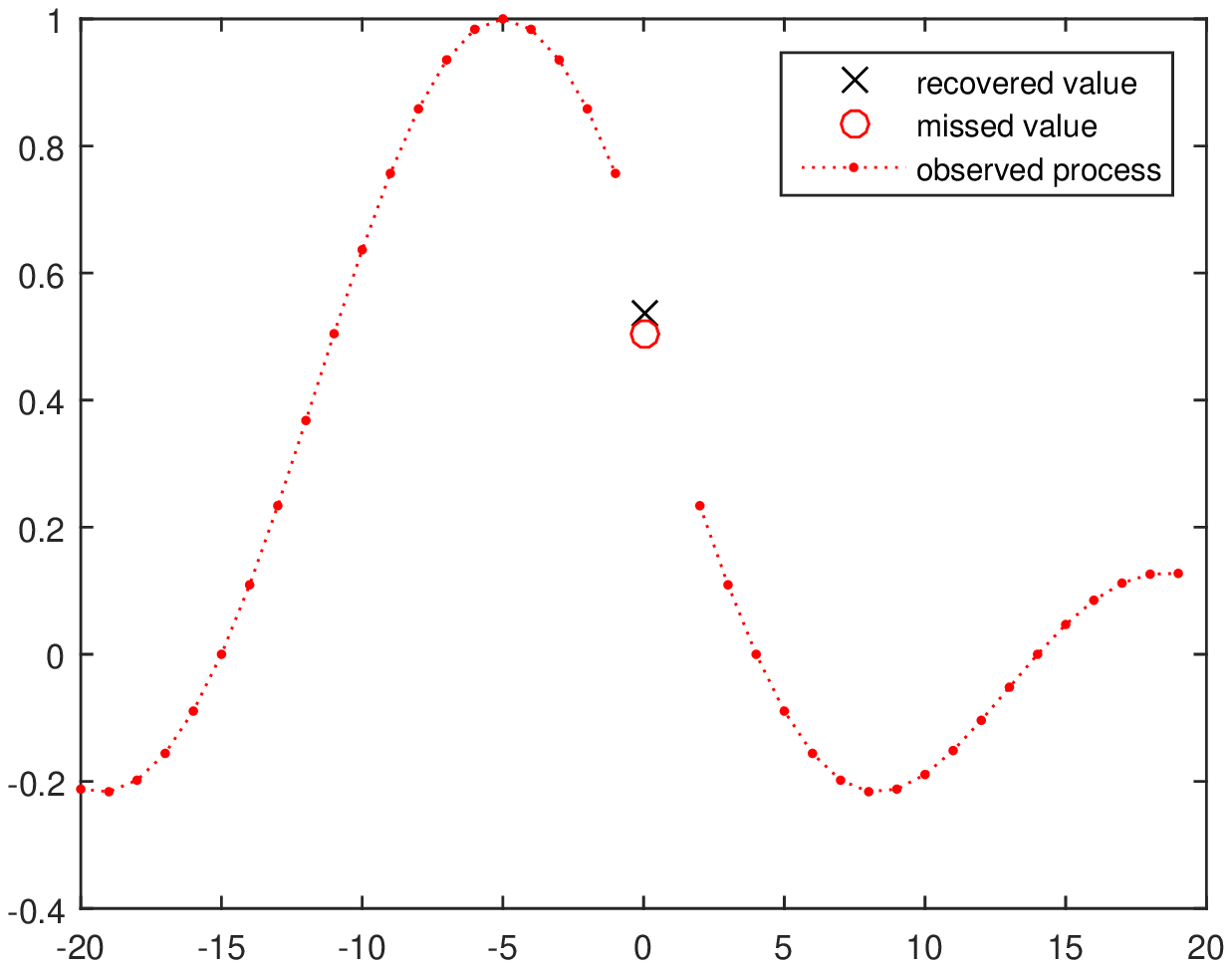,width=9cm,height=5.5cm}}
\caption[]{\sm
Example  of a path  $x\in\XN$ with $\O=0.1\pi$ and the recovered values   $\w x(0)$ calculated using 100 observations: (i) calculated by  (\ref{wx1})
for $\O=0.1\pi$  (top);   (ii) calculated by (\ref{wx1})  with $\O=0.05\pi$ (middle);
(iii) calculated by (\ref{wx2})  (bottom). } \vspace{0cm}\label{fig-1}
\end{figure}
\begin{figure}[ht]
\centerline{\psfig{figure=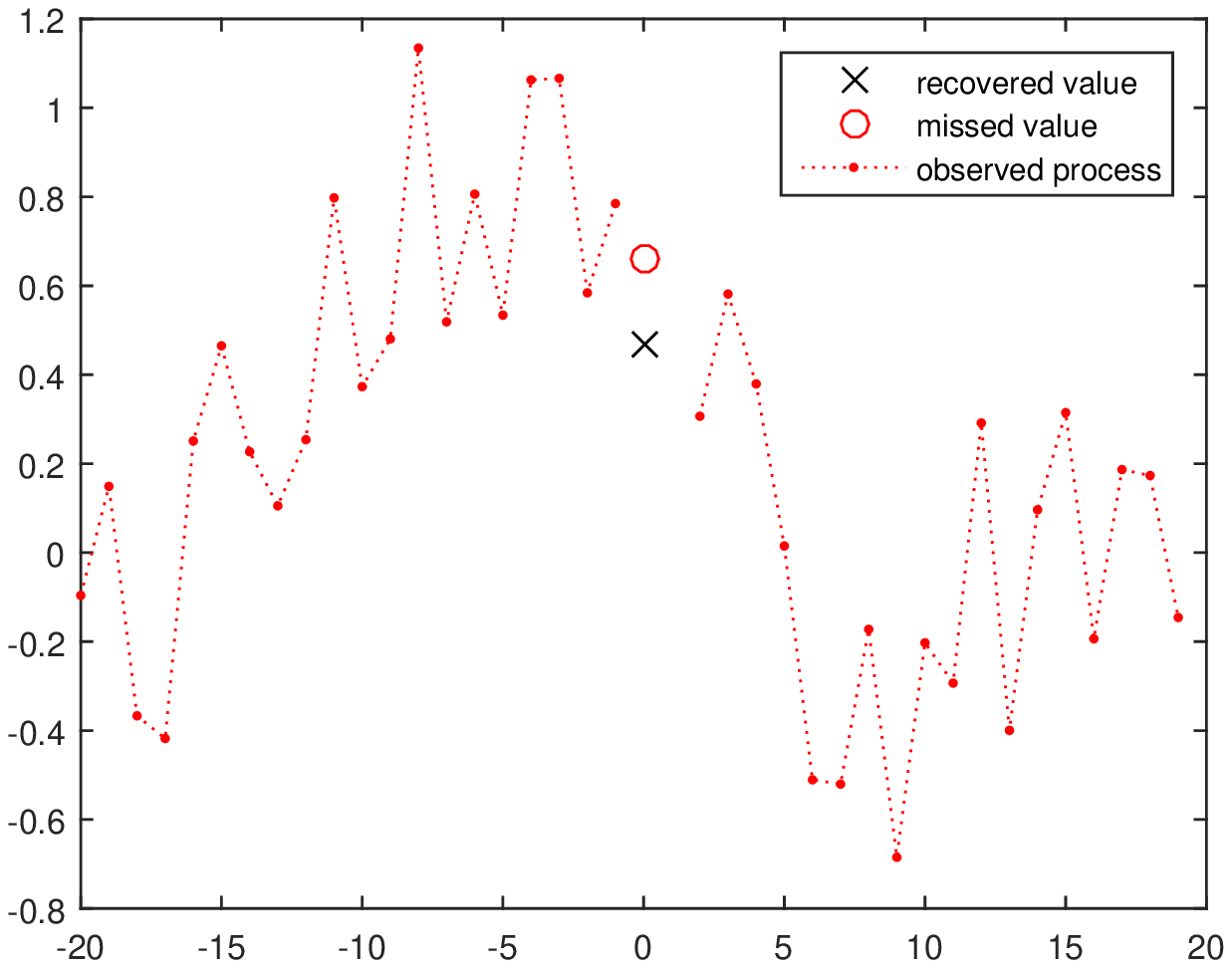,width=9cm,height=5.5cm}}
\centerline{\psfig{figure=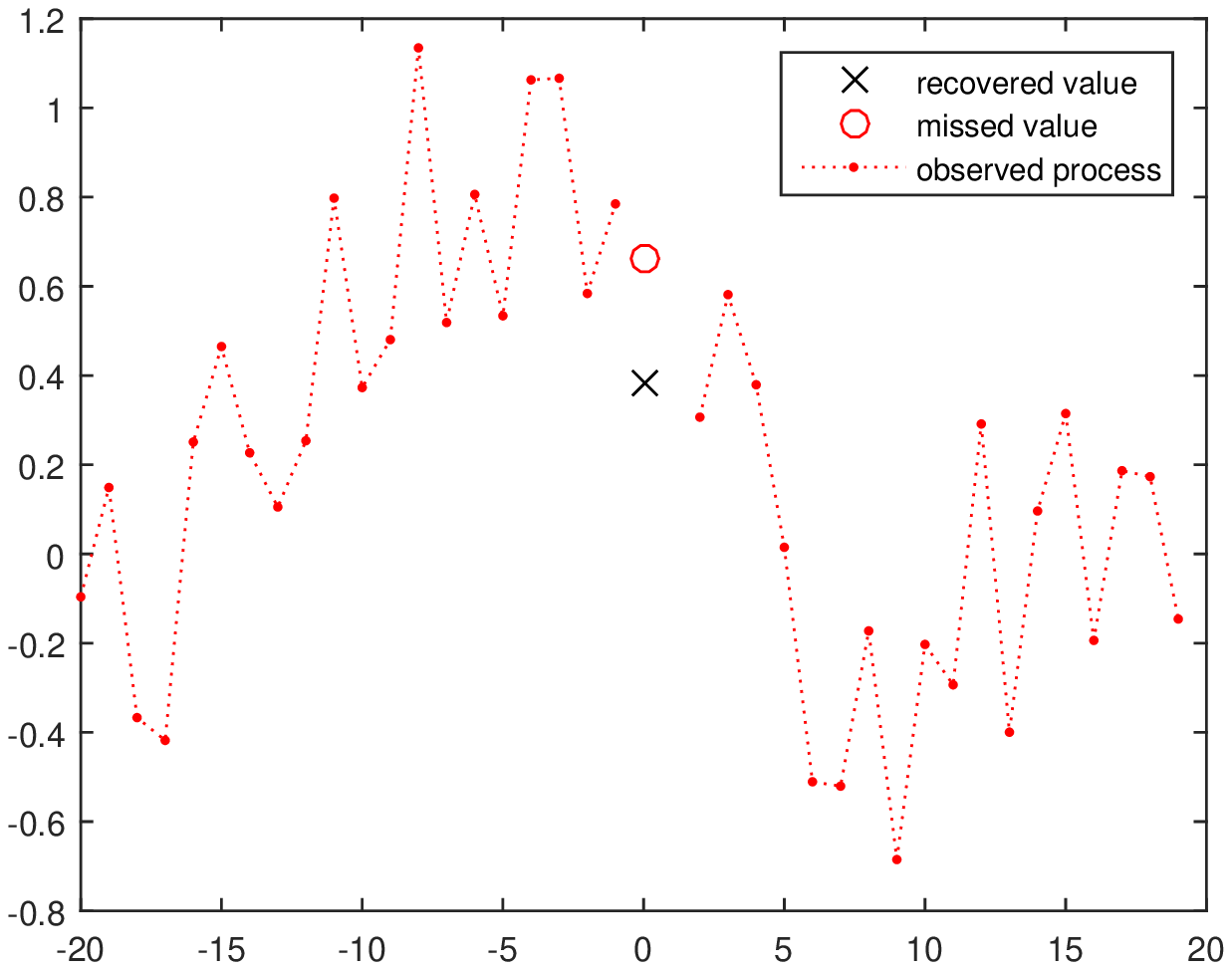,width=9cm,height=5.5cm}}
\centerline{\psfig{figure=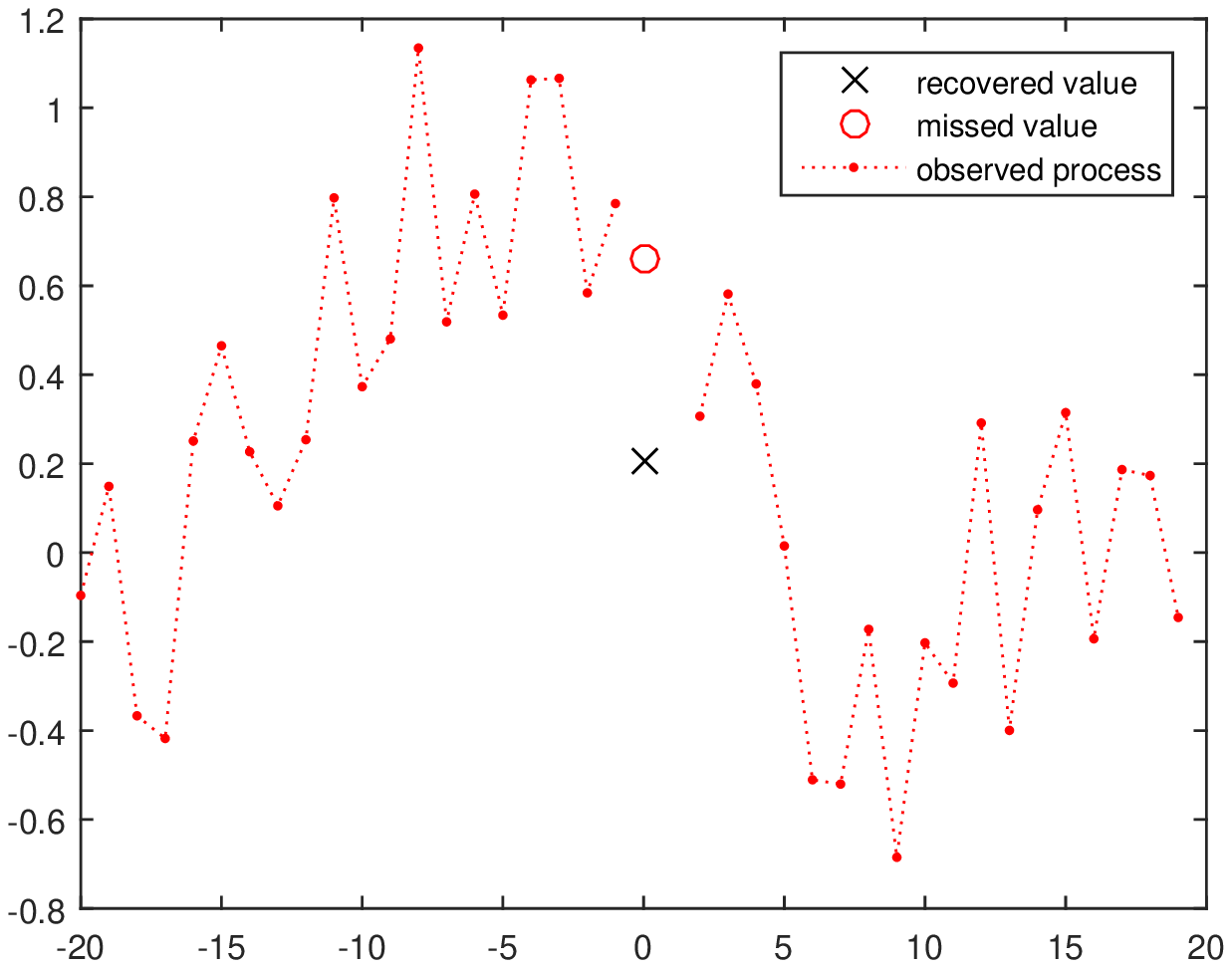,width=9cm,height=5.5cm}}
\caption[]{\sm
Example  of a path  $x\in\ell_2(\D)$ and the recovered values   $\w x(0)$ calculated using 100 observations: (i) calculated by  (\ref{wx1})
for $\O=0.1\pi$  (top);   (ii) calculated by (\ref{wx1})  with $\O=0.05\pi$ (middle);
(iii) calculated by (\ref{wx2})  (bottom). } \vspace{0cm}\label{fig-2}
\end{figure}

  On the other hand,   application of (\ref{wx2}), where $\O$ is not used,  performs better
   than (\ref{wx1}) with too small miscalculated $\O_1$.  This is illustrated by
Figure \ref{fig-1} that shows an  example of  a process $x(t)\in\XN$ with $\O=0.1\pi$ and recovered values $\w x(0)$
corresponding to band-limited extensions   obtained  from (\ref{wx1})
 with  $\O=0.1\pi$ and $\O=0.05\pi$. In addition, this figure shows $\w x(0)$
 calculated by (\ref{wx2}).
On the hand, the presence of  a noise in processes that are nor recoverable without error may lead to a larger
error for estimate (\ref{wx2}). This is illustrated by
Figure \ref{fig-2} that shows an  example of  a noisy process $x$ and recovered values $\w x(0)$
corresponding to band-limited extensions   obtained  from (\ref{wx1})
 with  $\O=0.1\pi$ and $\O=0.05\pi$. In addition, this figure shows $\w x(0)$
 calculated by (\ref{wx2}).
In these experiments, we used  $M_s=\{0\}$ and truncated sums (\ref{wx1}) and (\ref{wx2}) with  100 members.
\section{Proofs}

{\em Proof of Proposition \ref{propU}}. It is known \citet{F92,F94a,F94} that a continuous time bandlimited function can be recovered without error from an oversampling
sequence where a  finite number of sample values is unknown.
This implies that  if $x\in\XN $ is such that $x(t)=0$ for
$t\in\D$, then $x\equiv 0$. Then  the proof of Proposition \ref{propU} follows. $\Box$

 {\em
Proof  of Lemma \ref{lemma1}.}  It
suffices to  prove that $\XNL $ is a closed linear subspace of
$\ell_2(\D)$.  In this case,  there exists a unique projection $\w
x|_{\D}$ of $x|_{\D}$ on $\XNL$, and the proof will be completed.
\par

Let $\BN$ be the set of all mappings $X:\T\to\C$ such that $X\ew\in L_2(-\pi,\pi)$
and such that $X\ew =0$ for $|\o|>\O$ for $X=\ Zx$.

Consider the mapping $\zeta:\BN \to \XNL$ such that
\baaa x(t)=(\zeta (X))(t)=\frac{1}{2\pi}\int_{-\pi}^\pi
X\left(e^{i\o}\right) e^{i\o t}d\o, \quad t\in\TT.\eaaa
 It is a linear
continuous operator. By Proposition \ref{propU}, it is a bijection.
\par
 Since  the mapping $\zeta:\BN \to \XNL$ is continuous, it follows that
the inverse mapping $\zeta^{-1}: \XNL\to\BN$ is also
continuous; see e.g. Corollary in Ch.II.5 \citet{Yosida}, p. 77. Since the
set $\BN$ is a closed linear subspace of $L_2(-\pi,\pi)$, it
follows that $\XNL$ is a closed linear subspace of $\ell_2(\D)$.
Then a solution $\w x$  of problem (\ref{min})
is such that $\w x|_{D}$ is  a projection of $x|_{D}$ on $\XNL$ which is unique.
 Then the proof of  Lemma \ref{lemma1} follows.  $\Box$
\par
{\em Proof of Lemma \ref{lemma3}.} Let $\oo y=\{\oo y_k\}_{k=0}^m\in\C^{m+1}$ be arbitrarily selected such that $\|\oo y\|_{\ell_2}\neq 0$.
Let $y\in \ell_2(\ZZ)$ be such that $y|_{\D}=0$ and that $\oo y=y|_M$.  In this case,
$y\notin \XN$;  it follows, for instance, from Proposition \ref{propU}.  Let $Y=\Z y$. We have that $\Z(h\circ y)=H\ew Y\ew$.
Hence   $\|H\ew Y\ew\|_{L_2(-\pi,\pi)}< \|Y\ew\|_{L_2(-\pi,\pi)}$. This implies that  $\|h\circ y\|_{\ell_2}< \|y\|_{\ell_2}$.
Hence
\baaa
\|\A \oo y\|_{\ell_2}=\|\Ind_{\M}(h\circ y)\|_{\ell_2}\le \|h\circ y\|_{\ell_2}< \|y\|_{\ell_2}=\|\oo y\|_{\ell_2}.
\eaaa
Since  the space $\ell_2(M)$  is finite dimensional, it follows that $\|\A\|_{2,2}<1$.
Then the  statement of Lemma \ref{lemma3} follows.  $\Box$

\par
{\em Proof of Theorem \ref{Th1}}.
Assume that the input sequences $\{x(t)\}_{t\in \D}$ are extended on  $M_s$
such that $x|_{M_s}=\w x|_{M_s}$, where $\w x$ is the optimal process that exists according to Lemma \ref{lemma1}. Then   $\w x$
is a unique solution of  the minimization problem \baa &&\hbox{Minimize}\quad  \sum_{t\in\ZZ}|x_\BL(t)-x(t)|^2 \quad\breakk\hbox{over}\quad x_\BL\in \XN.\label{minP} \eaa

By the property of the low-pass filters, $\w x= h\circ x$.  Hence the optimal process $\w x\in \XN$ from Lemma \ref{lemma1} is such that
\baaa \w x=h\circ\left( x\Ind_{\D}+\w x\Ind_{M_s}\right). \eaaa
Hence
\baa \w x(t)=\sum_{s\in \D} h(t-s) x(s)+\sum_{s\in M_s} h(t-s) \w x(s).
\label{xy}\eaa
This gives that
\baaa
x(t)-\sum_{s\in M_s}\A_{t,s} x(s)=z_t.
\eaaa
This gives (\ref{yy1})-(\ref{yy3}).
 $\Box$
\par
{\em Proof of  Corollary
\ref{corr1}}. If $x\in \XN$, then $\w x=x$, since it is a solution of (\ref{min}).
 By Theorem \ref{Th1},  $\w x$ is obtained as is required in Definition \ref{def1} with $r=2$ and $\Y=\XN$. $\Box$
\par
{\em Proof of Lemma \ref{lemmaM}}.  The case where $m=0$ is trivial, since $\B(\o)= e^{-\o s}$ in this case.
 Let us consider the case where $m>0$;  by the assumptions, $s=0$ in this case.
 Suppose that there exists $\o\in(-\pi,\pi]$ such that the matrix $\B(\o)$ is degenerate. In this case, there   exists $q=\{q(k)\}_{k=0}^m\in \C^{m+1}$ such that
$q\neq 0$ and
$\B(\o)y=0$.  Let $Q(z)\defi \sum_{k=s}^{s+m} q(k)z^k=\sum_{k=0}^{m} q(k)z^k$, $z\in\C$. By the definition of $\B(\o)$, it follows that
$\frac{d^p Q}{d\o^p}\ew=0$ for $p=0,1,...,m$. Hence  $\frac{d^p Q}{dz^p}(z_0)=0$ at $z_0=e^{i\o}$ for $p=0,1,...,m$. Hence $Q\equiv 0$.  Therefore,
the vector $q$ cannot be non-zero. This completes the proof. $\Box$
\par
{\em Proof of Theorem \ref{Th2}}.
Let $y\in\ell_1$ be selected such that $y(t)=x(t)$ for $t\notin M_s$ and $y|_{M_s}=0$. Let $Y=\Z y$,
and let $\w x\in\ell_1$ be selected such that $\w x(t)=x(t)$  for $t\notin M_s$, with some choice of  $\w x|_{M_s}$.  Let $\w X=\Z \w x$.
It follows from the definitions that  \baaa
\frac{d^p\w X}{d\o^p}\ew=\frac{d^p Y}{d\o^p}\ew+\sum_{t=s}^{s+m}(-i\o t)^p e^{-i\o t}\w x(t)\brea=-z_p(\o)+\d_p(\B(\o)y(\o)), \quad p=0,1,...,m.
\eaaa

For $\o=\om$, this gives  $\B(\om)y(\om)=z(\om)$. Hence there is a unique choice  that ensures that $\w x\in \ellz$
and $\w x|_\D=x|_\D$; this choice is defined by equations
(\ref{y1})-(\ref{y3}). Clearly, this is a unique optimal solution of the minimization problem (\ref{opt}) with $r=1$ and $\Y=\X_0$.
This completes the proof of Theorem \ref{Th2}. $\Box$

{\em Proof of  Proposition  \ref{prop1}}. It suffices to prove statement (ii) only, since statment (i) is its special case.
Let $x\in\X_\s$ for some $\s\neq 0$, and let $Y\ew = \sum_{k\in\D} e^{-i\o k}x(k)$, $\o\in(-\pi,\pi]$; this function is
observable. By the definitions, it follows that
 \baaa
X\ew =Y\ew + \sum_{t\in M_s} e^{-i\o k}x(t)
 \eaaa
and \baaa
\frac{d^p X}{d\o^p}\ew =\frac{d^p Y}{d\o^p}\ew + \d_p(\B(\o)y(\o)),\quad p=0,1,...,m.
 \eaaa
 For $\o=\om$, it gives
 \baaa
\xi=-z(\om) + \B(\om)y(\om),
 \eaaa
where  $\xi=\{\xi_p\}_{p=0}^m\in\C^{m+1}$ has components
$\xi_p=\frac{d^p X}{d\o^p}\left(e^{i\om}\right)$
such that $|\xi_p|\le \s_p$. Using the estimator from Theorem \ref{Th2}, we accept the value
$\w y(\om)=\B(\om)^{-1}z(\om)$ as the estimate of $y(\om)=\{x(s+p)\}_{p=0}^m$.
We have that $\B(\om) y(\om)-\B(\om)\w y(\om)=\xi$.  It follows that  the first inequality in (\ref{opt}) holds.
If $\s=0$ then the estimator is error-free.
\par
Let us show that the second inequality in (\ref{opt}) holds.
Suppose that we use  another  estimator $\ww x(s)=\ww F\left(x|_\D\right)$, where $\ww F:\ell_2(\D)\to\C$ is some mapping.
Let $p\in\{0,1,...,m\}$, and let $X_\pm\ew$ be such that $\d_k(\B(\o)y(\o))=\pm \s_k\Ind_{\{k=p\}}$,  $k\in\{0,1,...,m\}$, and $x_\pm(t)=0$ for $t\in \D$ for
$x_\pm=\Z^{-1}X_\pm$. By the definition of $\B(\o)$, it follows  $\frac{d^k X_\pm}{d\o^k}\ew =\pm \s_k \Ind_{\{k=p\}}$.
Clearly, $x_\pm\in\X_\s$. Moreover, we have that $\ww x_-|_{M_s}=\ww x_+|_{M_s}$ for $\ww x_\pm= \ww F\left(x_\pm|_\D\right)$,
for any choice of $\ww F$, and \baaa
\max(|\d_p(\B(\om) \eta_-)|, |\d_p(\B(\om) \eta_+)|)\ge \s_p, \brea \quad  p=0,1,...,m,
\label{opttt}
\eaaa
where
$\eta_-=\{\ww x_-(t) -x_-(t)\}_{t=s}^{s+m}\in\C^{m+1}$,
$\eta_+=\{\ww x_+(t) -x_+(t)\}_{t=s}^{s+m}\in\C^{m+1}$.
 Then the second  inequality in (\ref{opt}) and the proof of  Proposition  \ref{prop1} follow. $\Box$
  \par
{\em Proof of Corollary
\ref{corr2}.} If $x\in \ellz$, then $\w x=x$ since it is a solution of (\ref{min}).
 By Theorem \ref{Th2},  $\w x$ is obtained as is required in Definition \ref{def1} with $r=1$ and $\Y=\ellz$. $\Box$
\par
{\em Proof of Proposition \ref{propR1}}.  By Theorem \ref{Th1},
\baaa
\|\w x|_{M_s}\|_{\ell_\t(M_s)}\le  \|(I_{m+1}-\A)^{-1}\|_{2,\t} \|z\|_{\ell_2(M_s)}.
\label{est1}
\eaaa
In addition,
\baaa
 \|z\|_{\ell_2(M_s)} \le \|\Ind_{M_s}(h\circ x\Ind_{\D})\|_{\ell_2(\ZZ)}
 \le  \|x|_{\D}\|_{\ell_2(\D)}.
\eaaa
Then the proof of Proposition \ref{propR1} follows. $\Box$
\par
{\em Proof of Proposition \ref{propR2}}.  By Theorem \ref{Th2},
\baaa
\|\w x|_{M_s}\|_{\ell_\t(M_s)}  \le  \|\B(\om)^{-1}\|_{\rho,\t} \|z(\om)\|_{\ell_\rho(\D)}.
\label{est21}
\eaaa
Further,
\baaa
|z_p(\om)|\le \sum_{t\in \D} |t|^m |x(t)|.
\label{est22}
\eaaa
Then the proof of Proposition \ref{propR1} follows. $\Box$
\section{Discussion and possible modifications}
The present paper is focused on theoretical aspects of possibility to recover
missing  values. The  paper suggests  frequency  criteria  of  error-free recoverability of  a single missing value
in pathwise deterministic setting.  In particular, $m$ missing values can be recovered for processes
that are  degenerate of order $m$ (Definition \ref{defD}).
Corollary \ref{corr2} gives a recoverability  criterion reminding  the classical Kolmogorov's  criterion (\ref{Km})  for the spectral densities \citet{K}.
However, the degree of similarity is quite limited. For instance, if a stationary Gaussian process has the
spectral density $\phi(\o)\ge \const\cdot (\pi^2-\o^2)^{\nu}$  for $\nu\in (0,1)$, then, according to criterion (\ref{Km}), this process is not minimal \citet{K},
i.e. this process is non-recoverable. On the other hand,  Corollary \ref{corr2}
imply that  single values of processes $x\in\ell_1$ are recoverable if $X(-1)=0$ for $X=\Z x$.  In
particular, this class includes sequences $x$ such that $|X\ew|\le \const \cdot (\pi^2-\o^2)^{\nu}$ for $\nu\in (0,1)$.
Nevertheless, this  similarity still
could be used for analysis of the properties  of pathwise Z-transforms for stochastic Gaussian processes.
In particular, assume that $y=\{y(t)\}_{t\in\ZZ}$
is a stochastic stationary Gaussian process with spectral density $\phi$ such that (\ref{Km}) does not hold. It follows that adjusted paths $\{(1+\d t^2)^{-1}y(t)\}_{t\in\ZZ}$, where $\d>0$, cannot belong to
 $\XN$ or $\X_0$.
We leave this analysis for the future research.


There are some other open questions. The most challenging problem is to obtain
 pathwise necessary  conditions of recoverability that are close enough to  sufficient conditions.
 In addition, there are more technical questions. In particular, it is unclear if it possible to
relax  conditions of recoverability described as weighted $\ell_1$-summarability
presented in the definition for $\X_\s$. It is also unclear if it is possible to replace
the restrictions on the derivatives of Z-transform imposed at one common point  for the processes from $\X_0$ by  conditions at different points.
 We leave this for the future research.

\index{this would require a setting closed to \cite{CTao}
with multiple restrictions for Z-transform, instead of the single restriction $\X(-1)=1$ imposed for the class $\elll$ in  Theorem \ref{Th2}.}

\subsection*{Acknowledgment}
This work  was supported by ARC grant of Australia DP120100928 to the author. In addition, the author thanks
the anonymous referees for their valuable suggestions  which
helped to improve the paper.

\end{document}